\documentclass{aa}

\usepackage[varg]{txfonts}
\usepackage{graphicx}
\usepackage{amsmath}
\usepackage[colorlinks=true,linkcolor=magenta,citecolor=blue,filecolor=cyan,urlcolor=black,final=true]{hyperref}

\makeatletter
\renewcommand*\aa@pageof{, page \thepage{} of \pageref*{LastPage}}
\makeatother
 
\begin{document} 

   \title{Three-dimensional reconstruction of multiple particle acceleration regions during a coronal mass ejection}
	\titlerunning{3D reconstruction of particle acceleration regions during a CME}

   \author{D. E. Morosan \inst{1}
        \and
        E. Palmerio \inst{1,2} 
        \and
        J. Pomoell \inst{1}
        \and
        R. Vainio \inst{3}
          \and
          M. Palmroth \inst{1,4}
          \and
          E. K. J. Kilpua \inst{1} 
          }

   \institute{Department of Physics, University of Helsinki, P.O. Box 64, FI-00014 Helsinki, Finland \\
              \email{diana.morosan@helsinki.fi}
         \and
            Space Sciences Laboratory, University of California--Berkeley, Berkeley, CA 94720, USA
         \and
             Department of Physics and Astronomy, University of Turku, FI-20014 Turku, Finland
        \and
            Finnish Meteorological Institute, Space and Earth Observation Centre, P.O. Box 503, FI-00014 Helsinki, Finland
             }

   \date{Received ; accepted }

 
  \abstract
    {Some of the most prominent sources for particle acceleration in our Solar System are large eruptions of magnetised plasma from the Sun called coronal mass ejections (CMEs). These accelerated particles can generate radio emission through various mechanisms.}
    {CMEs are often accompanied by a variety of solar radio bursts with different shapes and characteristics in dynamic spectra. Radio bursts directly associated with CMEs often show movement in the direction of CME expansion. Here, we aim to determine the emission mechanism of multiple moving radio bursts that accompanied a flare and CME that took place on 14 June 2012.}
    {We used radio imaging from the Nan{\c c}ay Radioheliograph, combined with observations from the Solar Dynamics Observatory and Solar Terrestrial Relations Observatory spacecraft, to analyse these moving radio bursts in order to determine their emission mechanism and three-dimensional (3D) location with respect to the expanding CME.}
    {In using a 3D representation of the particle acceleration locations in relation to the overlying coronal magnetic field and the CME propagation, for the first time, we provide evidence that these moving radio bursts originate near the CME flanks and some that are possible signatures of shock-accelerated electrons following the fast CME expansion in the low corona.}
    {The moving radio bursts, as well as other stationary bursts observed during the eruption, occur simultaneously with a type IV continuum in dynamic spectra, which is not usually associated with emission at the CME flanks. Our results show that moving radio bursts that could traditionally be classified as moving type IVs can represent shock signatures associated with CME flanks or plasma emission inside the CME behind its flanks, which are closely related to the lateral expansion of the CME in the low corona. In addition, the acceleration of electrons generating this radio emission appears to be favoured at the CME flanks, where the CME encounters coronal streamers and open field regions.}


\maketitle


\section{Introduction}

{Flares and coronal mass ejections (CMEs) from the Sun are the most powerful and spectacular explosions in the Solar System that are capable of releasing vast amounts of magnetic energy over a relatively short period of time. These phenomena are often associated with particle acceleration processes, including acceleration of electrons that can generate emission at radio wavelengths through various mechanisms.}

{The standard picture of CMEs predicts that they generally erupt when twisted or when sheared magnetic fields in the corona become unstable \citep{ch17,gr18}. As they propagate to higher coronal heights, CMEs are often seen in white-light imagery as a bright core and a dark cavity surrounded by a bright compression front \citep[i.e. the classic three-part structure;][]{il85}. The cavity is believed to correspond to a magnetic helical structure, known as a `flux rope' \citep{de99,vo13}, which is an integral part of every CME regardless of its pre-eruptive configuration \citep{fo00,au10,ch11}. When CMEs propagate faster than the characteristic speed of the ambient medium, they can drive shock waves that are capable of accelerating electrons to high energies. CME-driven shocks can be observed in white light as a fainter bright emission that surrounds the CME bubble \citep{vo03,vo13}. Classically, the most obvious manifestations of shocks at radio wavelengths are a class of radio bursts called type II bursts, which are emitted by the plasma emission mechanism \citep{ne85,ma96,kl02}. Type II radio bursts show emission sources that are closely associated with expanding CME-driven shocks in the corona \citep{zu18, st74}. Fine-structured bursts called herringbones, which can accompany type II bursts or occur on their own \citep{ho83, ca87, ca89}, represent signatures of individual electron beams accelerated by a shock, predominantly at the CME flanks \citep{ca13, mo19a}.  }

\begin{figure*}[ht]
    \centering
    \includegraphics[width=0.7\linewidth]{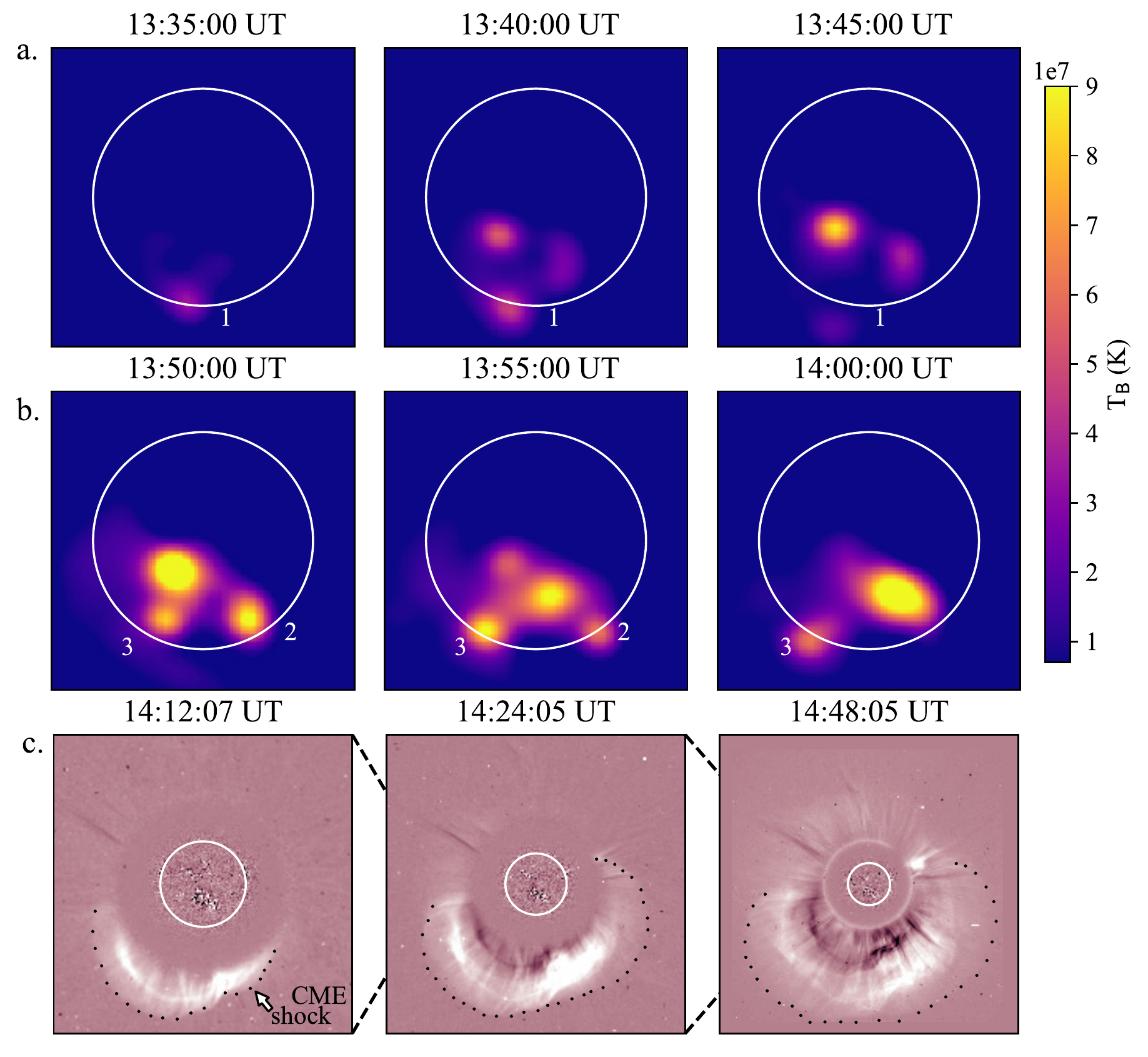}
    \caption{Moving radio bursts associated with the expansion of the 14~June~2012 CME in the solar corona. The radio bursts are shown in panels (a) and (b) from 13:35 to 14:00~UT as observed by the NRH at 150~MHz. The white-light signature of the CME becomes visible at a later time (from 14:12~UT), when it expands to the higher corona as observed by the LASCO/C2 coronagraph. The white circles in the images in panel (c) denote the visible solar limb. The white arrow points towards the CME shock, which is indicated by the fainter outline surrounding the CME. The outer edge of the white-light shock is denoted by the dotted line in the panel (c). A clear shock is observed in both south-easterly and south-westerly directions.}
    \label{fig:fig1}
\end{figure*}

{CMEs can also be accompanied by continuum emission at decimetric and metric wavelengths, known as type IV bursts, that can have either stationary or moving sources \citep[for a review, see][]{ba98}. Moving type IV radio bursts, in particular, are commonly believed to be generated either by electrons trapped inside the CME flux rope that emit gyro-synchrotron radiation \citep{bo68, du73} or electrons accelerated at a reconnecting current sheet that form at the wake of an erupting flux rope, generating radio bursts at the plasma frequency \citep{ga85, vr03, mo19b}. In the former case, the radio source thus moves with the outward propagating CME and in the latter case the current sheet producing the type IV electrons also moves upwards in the corona as it is dragged by the rising CME. Moving type IV radio bursts, as well as type IIs and herringbones, can therefore be directly related to the outward propagation of CMEs in the solar corona. However, only type II radio bursts and herringbones have been directly related to CME shocks. So far, electron acceleration has only been observed at selective locations at the CME flanks in the low corona \citep{ca13,mo19a}, despite the extended structure of the shock. The three-dimensional (3D) location of electron acceleration regions in relation to the propagating CME structure and surrounding corona is also not well known. }

{In this paper, we present observations of the radio emission associated with a CME that erupted on 14 June 2012. Using 3D modelling of the magnetic field environment and the CME structure, we show that multiple moving radio sources accompanying the CME represent plasma radiation near the CME flanks, where the CME is most likely to drive a shock. In Sect.~\ref{sec:analysis}, we give an overview of the observations and data analysis techniques used. In Sect.~\ref{sec:results}, we introduce the results of the moving radio bursts analysis, which are further discussed in Sect.~\ref{sec:discussion}. Finally, we present our conclusions in Sect.~\ref{sec:conclusions}.}

\begin{figure*}[ht]
\centering
    \includegraphics[width=0.8\linewidth]{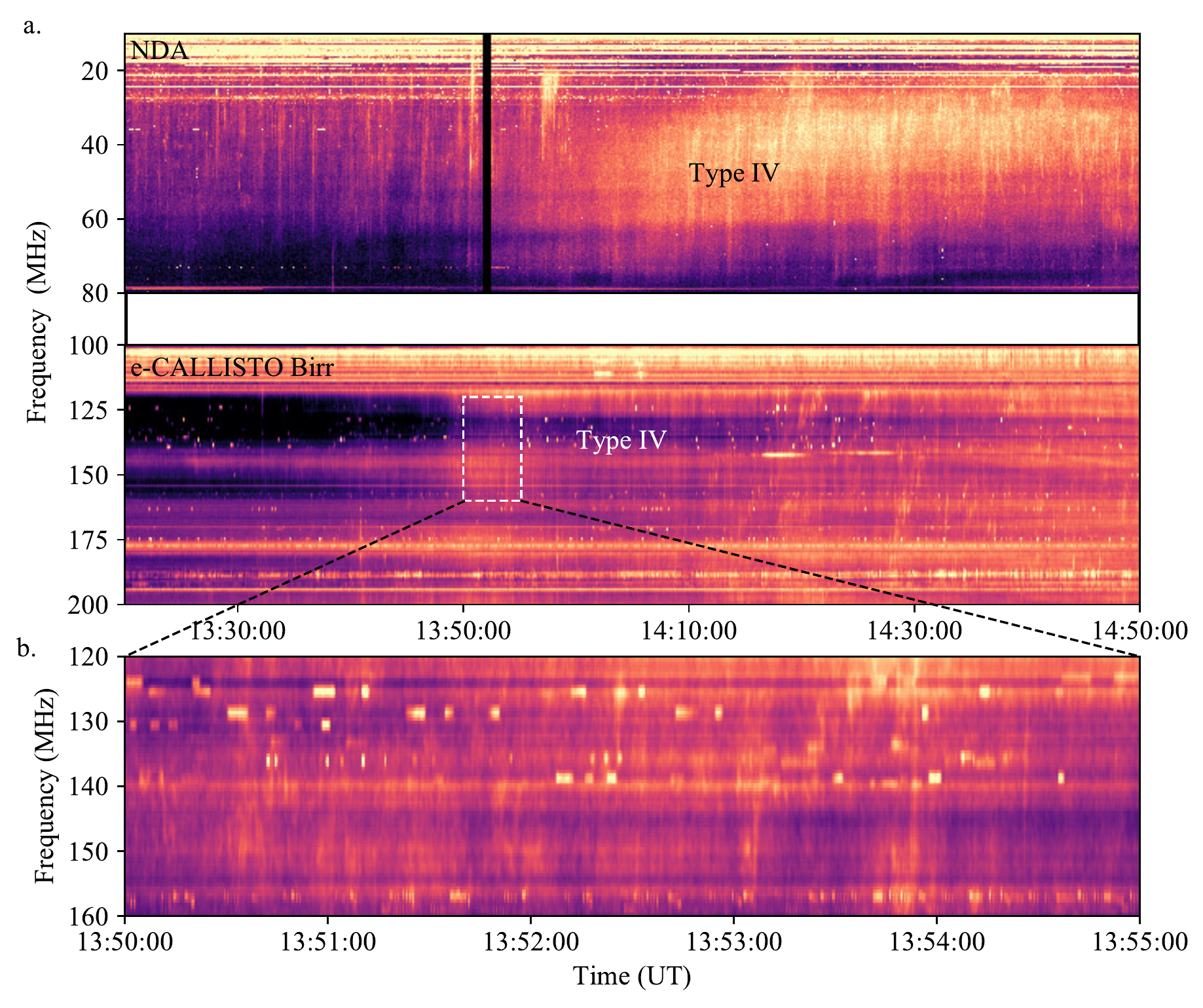}
    \caption{Dynamic spectrum of the type IV radio burst observed on 14 June 2012. (a) Dynamic spectra from the NDA (10--80~MHz) and the e-CALLISTO Birr spectrometer (100--200~MHz) showing a type IV continuum. (b) Zoom-in of the dynamic spectrum in (a) showing some of the fine scale structures inside the type IV continuum at NRH frequencies.}
    \label{fig:fig2}
\end{figure*}


\section{Observations and data analysis} \label{sec:analysis}

{On 14 June 2012, three moving radio sources were observed in images from the Nan{\c c}ay Radioheliograph \citep[NRH;][]{ke97} that appeared to follow the propagation direction of a CME accompanying a M1.9-class solar flare. This CME was well observed in remote-sensing observations by the Solar and Heliospheric Observatory \citep[SOHO;][]{do95} and Solar Dynamics Observatory \citep[SDO;][]{pe12} spacecraft located near Earth. The CME was also observed by the twin Solar Terrestrial Relations Observatory \citep[STEREO;][]{ka08} spacecraft orbiting the Sun close to 1~AU and, at the time of this study, separated from Earth by $116^{\circ}$ and $117^{\circ}$, respectively. The onset and eruption of the 14~June~2012 CME from multi-wavelength imagery of the solar disc have been analysed extensively in several studies \citep{ja17,pa17,wa19}. The first observation of the CME in white-light coronagraph images was at 13:25~UT in the STEREO’s inner coronagraph COR1, which is part of the Sun Earth Connection Coronal and Heliospheric Investigation \citep[SECCHI;][]{ho08} suite.}

{The NRH radio data\footnote{\url{https://rsdb.obs-nancay.fr}} were processed using the standard SolarSoft NRH package to produce calibrated radio images of the Sun. In the NRH images, we identified three radio sources moving away from the Sun in south-westerly and south-easterly directions and that are labelled 1, 2, and 3 in order of appearance in Fig.~\ref{fig:fig1}a--b. Multiple other stationary radio sources were also observed associated with the CME eruption (see e.g. the unlabelled radio sources in Fig.~\ref{fig:fig1}.) The bottom row in Fig.~\ref{fig:fig1} shows the CME ${\sim}12$ minutes later in images from the Large Angle Spectrographic Coronagraph \citep[LASCO;][]{br95} onboard SOHO. The CME propagated away from the Sun and towards the observer from Earth's perspective (Fig.~\ref{fig:fig1}c). The CME was relatively fast, with a plane-of-sky speed of 987~km/s as reported in the CDAW SOHO/LASCO CME Catalog\footnote{\url{https://cdaw.gsfc.nasa.gov/CME_list/}}, and it expanded rapidly in longitude. A CME-driven shock is also visible in Fig.~\ref{fig:fig1}c (outlined by the black dots in all panels), identified as a sharp, fainter outer boundary surrounding the CME.}

{In dynamic spectra from the Nan{\c c}ay Decametric Array \citep[NDA;][]{bo80} and the e-Callisto Birr spectrometer \citep{zu12}, a period of continuum radio emission was observed that extended in frequency from a few hundred MHz to 20~MHz and lasted for ${\sim}2$~hours, starting at 13:30~UT (Fig.~\ref{fig:fig2}a). This emission occurred approximately at the same time as the apex of the white-light front of the CME reached a height of $1.5\,R_{\odot}$ (i.e. the inner edge of the STEREO/COR1 field of view). Such continuum radio emission is identified as a type IV radio burst, although it can consist of multiple components with varying emission mechanisms \citep[i.e. plasma emission or gyro-synchrotron emission;][]{mo19b}. Unfortunately, the resolution and sensitivity of the e-CALLISTO spectrometer, which observed the type IV burst at the same frequencies as NRH, are low. Therefore, it is hard to distinguish different components in the type IV continuum, however some fine-structured bursts are superimposed on this emission but they are not clearly identifiable  (Fig.\ref{fig:fig2}b). These fine structured bursts resemble herringbone bursts and, if herringbones or similar bursts are indeed present inside the type IV continuum, they would indicate the presence of electrons accelerated by a CME shock \citep{ca87, mo19a}. Previous studies by \citet{ca13} have already identified herringbone bursts in the radio continuum following a type II burst, however in our case we do not see any clear signatures of shock-accelerated electrons in the dynamic spectra, such as a type II or herringbones.}

\begin{figure*}[ht]
\centering
    \includegraphics[width=0.85\linewidth]{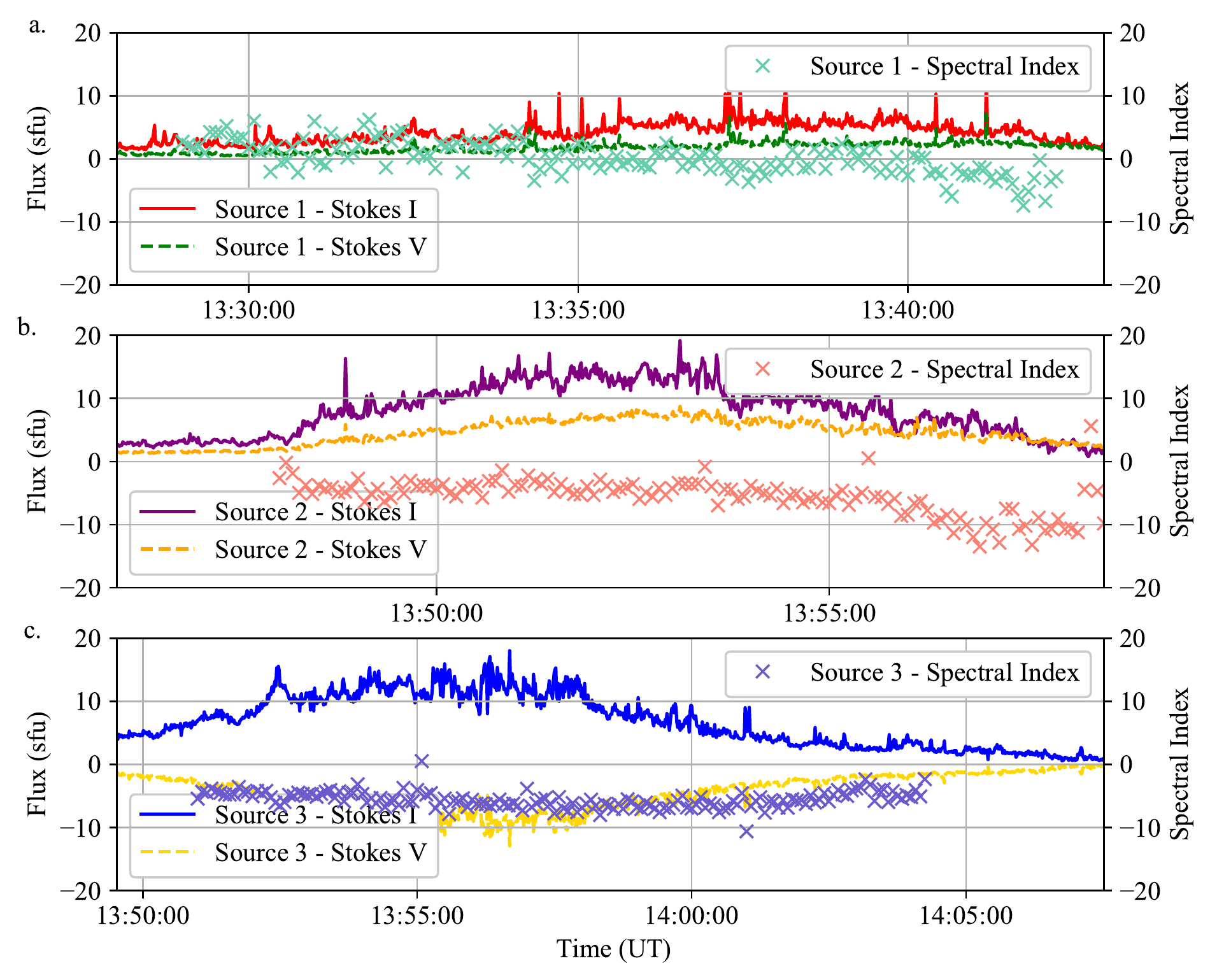}
    \caption{Flux densities and spectral indices of the moving radio sources 1 (a), 2 (b), and 3 (c).  The flux densities of the moving radio sources are estimated for both Stokes I (solid lines) and Stokes V (dashed lines) emission in each panel. The spectral indices were estimated by fitting a power-law function to the lower NRH frequencies, where the radio sources are clearly observed in images: 150--228~MHz, 150--170~MHz, and 150--170~MHz, for Sources 1, 2, and 3, respectively.}
    \label{fig:fig3}
\end{figure*}

\begin{figure*}[ht]
\centering
    \includegraphics[width=0.8\linewidth]{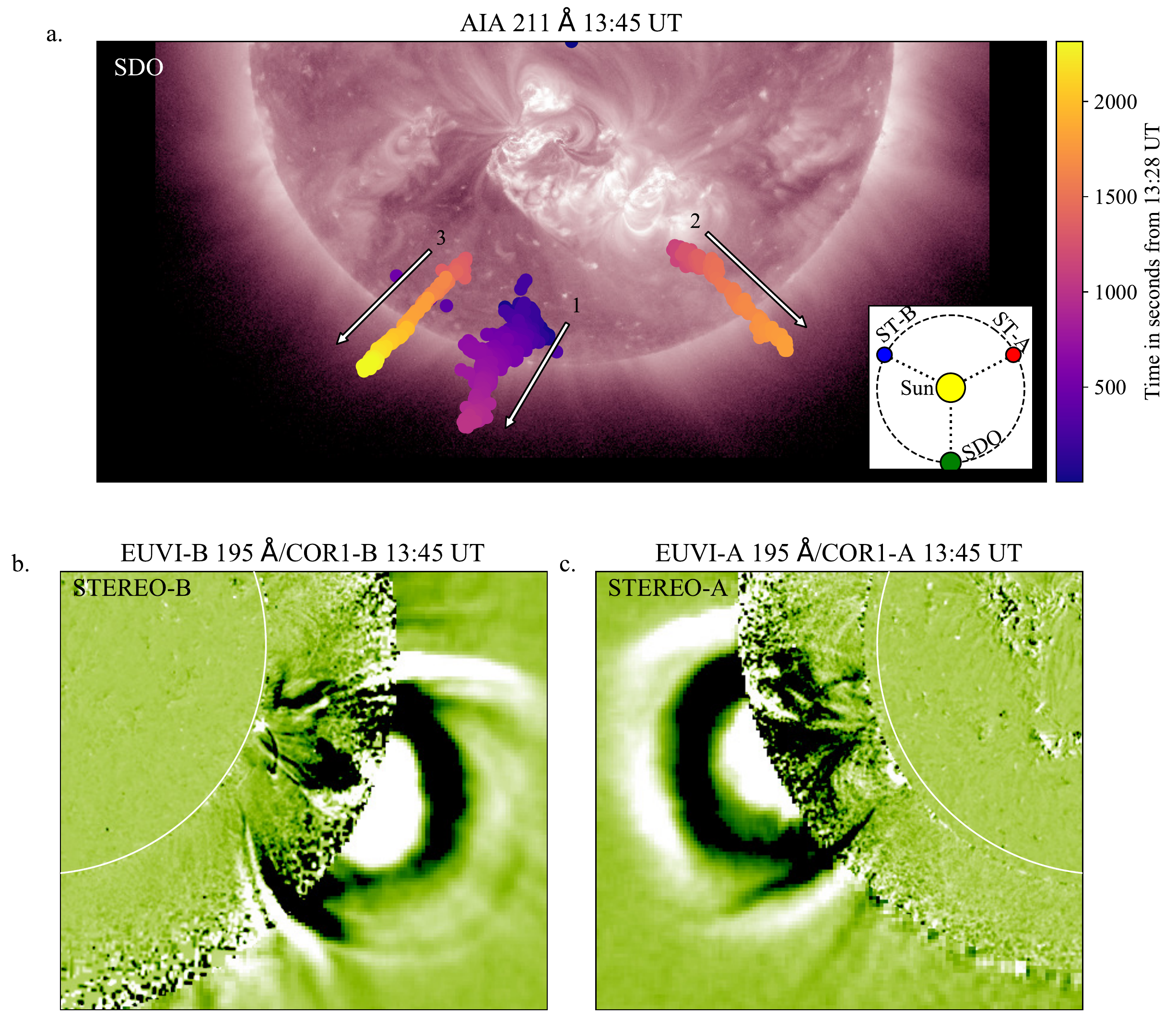}
    \caption{Centroids of moving radio sources through time and associated solar activity visible from three perspectives. (a) Centroids of the three moving sources identified in Fig.~\ref{fig:fig1}, overlaid on an SDO/AIA 211~{\AA} image of the Sun at 13:45~UT. The centroids are colour-coded through time from 13:28 until 14:06~UT. In the AIA image, the source active region appears bright as a result of the sudden energy release of the M-class flare, however the accompanying CME is not visible. (b) Solar activity as viewed from the STEREO-B perspective that captures a side view of the flare and associated CME from the eastern side with respect to the SDO field of view at 13:45~UT. (c) Solar activity as viewed from the STEREO-A perspective that captures a side view of the flare and associated CME from the western side with respect to the SDO field of view at 13:45~UT. The EUV images at 195~{\AA} from EUVI are combined with coronograph images from the COR1 cameras onboard both STEREO-A and -B. The locations of SDO, STEREO-A, and STEREO-B in the ecliptic plane relative to the Sun can be seen in the insert in panel (a). }
    \label{fig:fig4}
\end{figure*}


\section{Results} \label{sec:results}

{The moving radio bursts in Fig.~\ref{fig:fig1} are mostly observed in images at 150 and 173~MHz from the NRH (see Movies~1 and 2 accompanying this paper), except for Source 1 that is also sometimes prominent at 228~MHz. This indicates that the moving radio sources are narrow-band emission corresponding to a narrow frequency region of the time period observed in the dynamic spectrum in Fig.~\ref{fig:fig2}a, below 200~MHz. It is important to note that higher frequency emission (>200~MHz), as well as other radio sources at 150 and 173~MHz, were also observed in NRH images. This emission can also be identified as type IV emission. However, this high-frequency radio emission seems to have no relation to the moving radio sources at 150 and 173~MHz, but it appears to correspond instead to stationary radio bursts located on top of the flaring active region during the CME expansion (see Fig.~\ref{fig:fig1}a--b and Movie~1 accompanying this paper). Here we only focus on the moving radio sources, while the stationary radio sources will be investigated separately in a follow-up study. We therefore refer to the radio sources that show movement in Fig.~\ref{fig:fig1} as moving radio bursts throughout this paper, as it is unclear if they are indeed part of the type IV continuum or if they are individual bursts superimposed on the type IV continuum.}

{The emission mechanism of the moving radio bursts can provide some insight into the origin of electron acceleration and their relation to the accompanying CME. Hence, the flux densities of the moving radio bursts are estimated for each of the three sources observed. These flux densities are estimated inside a box covering the full extent and movement of the radio source and they include the pixels with levels >20\% of the maximum intensity levels in each box. We choose a treshold of 20\% to include the full extent of the radio source and exclude quiet-Sun emission. The flux densities are estimated in solar flux units (sfu; where 1~sfu = $10^{-22}$~W~m$^{-2}$~Hz$^{-1}$) for total intensity (Stokes I) and intensity of circularly polarised radiation (Stokes V) and are presented in Fig.~\ref{fig:fig3}. We note that there is an uncertainty in the absolute calibration of NRH data of ${\sim}$10--20\% \citep[see e.g.][]{dr99,ca17}. The emission is bursty and highly circularly polarised in the case of all moving radio sources. We also estimate the spectral index of the emission, $\alpha$, to help determine the possible emission mechanism. This is done by fitting a power-law function to NRH flux density spectra (plots of flux density vs. frequency), where the spectral index $\alpha$ is defined as the slope of a log--log plot:
\begin{equation}
\alpha = \frac{d \log S}{d \log \nu} \, ,
\end{equation}
where $S$ is the flux and $\nu$ is the observed frequency. The spectral index is fitted to two to three data points, since the moving radio bursts were only observed at 150 and 170~MHz and rarely at 228~MHz. There were no moving radio bursts observed at frequencies above 228~MHz. The spectral index evolution through time is denoted by the $\times$-symbols in Fig.~\ref{fig:fig3}.}

\begin{figure*}[ht]
\centering
    \includegraphics[width=0.85\linewidth]{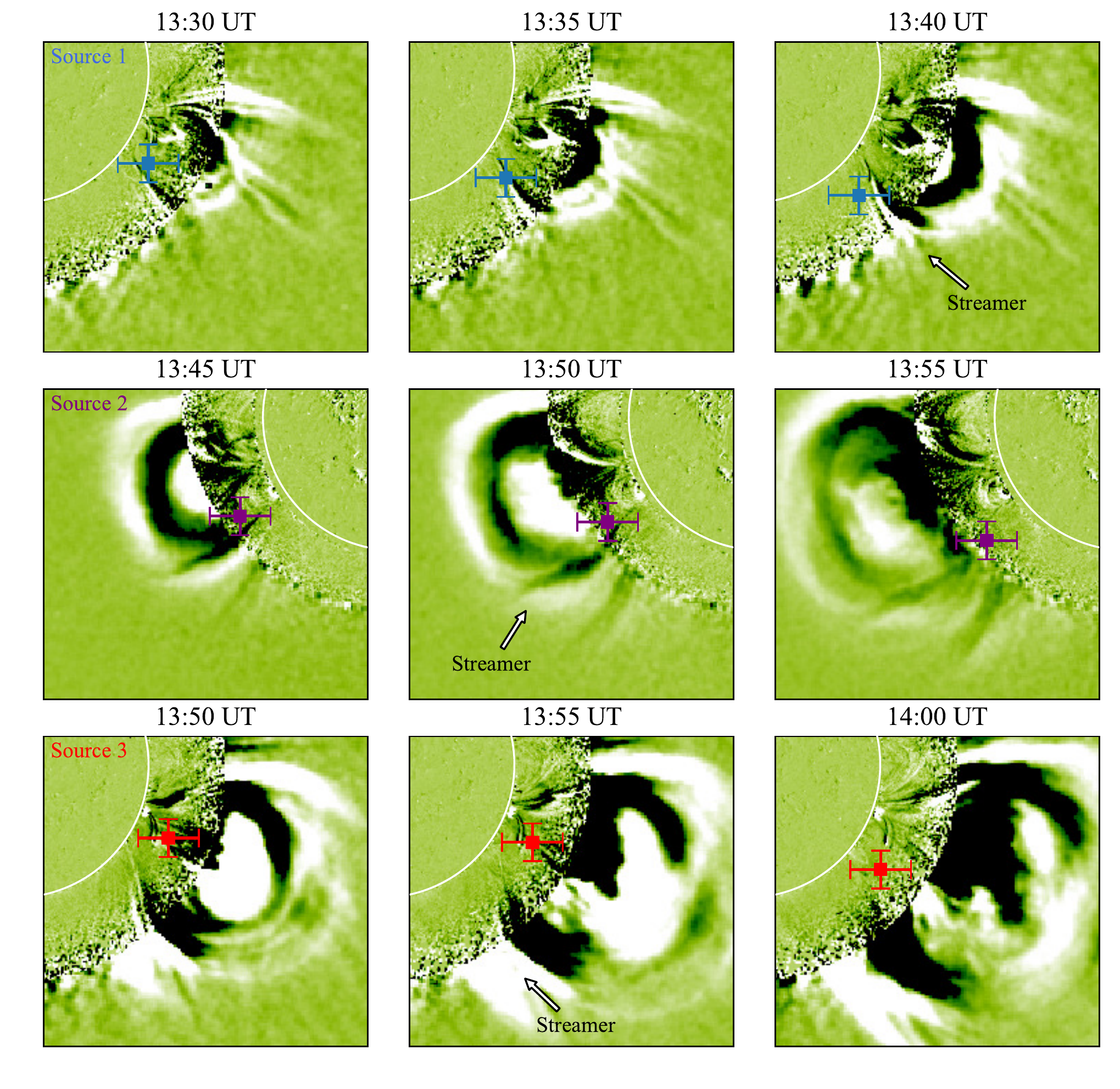}
    \caption{Centroids of the moving radio sources projected onto the STEREO-A and -B perspectives over time. The STEREO images in each panel are composed of EUV images at 195~{\AA} from EUVI and coronagraph images from the COR-1 instruments onboard both STEREO-A and B. The top and bottom panels show the STEREO-B perspective, where the first and third moving sources can be seen, while the middle panel shows the STEREO-A perspective, where the second moving source can be observed. The radio centroids coincide with the southern flank of the CME moving further southwards in each panel. }
    \label{fig:fig5}
\end{figure*}

{All moving radio sources have a high degree of circular polarisation (see Fig.~\ref{fig:fig3} and Movie 2 accompanying this paper). The first moving source (Source 1) consists of bursty emission that is 30--60\% positively circularly polarised based on our flux estimates and over the time interval shown. Source 1 is less bright compared to the later moving sources (Fig.~\ref{fig:fig3}a). The second and third moving sources (Sources 2 and 3) also consist of bursty emission that is up to 90\% circularly polarised (Fig.~\ref{fig:fig3}b--c), based on our flux estimates. We note, however, that the exact percentage of circular polarisation is difficult to estimate due to possible errors in the NRH absolute calibration procedures. However, the high signal-to-noise ratio in the Stokes V images in the case of the moving radio sources indicates that there is indeed a strong component of circular polarisation. Sources 2 and 3 have an opposite sense of circular polarisation. Source 2 is positively polarised, similarly to Source 1, while Source 3 is negatively polarised. This indicates different environments for the propagation of emitting electrons, such as access to differently orientated magnetic field lines (see Movie~2 accompanying this paper for the full evolution of the polarised moving sources). The spectral indices of all three moving sources are negative and steep, especially for Sources 2 and 3 that have spectral indices below $-4$ throughout their duration (see Fig.~\ref{fig:fig3}). In previous studies, the first moving burst (Source 1) was designated as emission originating from electrons trapped within the CME accelerated along magnetic field lines and emitting gyro-synchrotron radiation \citep{ja17}. However, the high degree of circular polarisation and steep spectral indices, especially for Sources 2 and 3 (Fig.~\ref{fig:fig3}b--c), are outside the range of known spectral indices for gyro-synchrotron emission. This suggests that the emission does not necessarily originate within the CME itself. Furthermore, the emission of all three moving sources is narrow-band and bursty in nature. These properties are all indicative of fundamental plasma emission \citep{ro78,be76}. Thus, in the next sections, we assume that the three moving radio sources are emitted by the plasma emission mechanism.}


\subsection{Radio source propagation from different perspectives}

{Since our analysis in the previous section indicates that the moving radio sources are unlikely to be gyro-synchrotron rdiation, here, we determine where the sources are located in relation to the CME body. The moving radio sources and CME originated within the visible solar disc (close to the central meridian), moving away from the Sun and towards the observer from Earth's perspective (Fig.~\ref{fig:fig4}). The radio centroids, coloured with progressing time, show the propagation direction of the radio bursts superposed on a plane-of-sky image at extreme ultraviolet (EUV) wavelengths from the Atmospheric Imaging Assembly \cite[AIA;][]{le12} onboard SDO (Fig.~\ref{fig:fig4}a). Source 1 and Source 3 move in a south-easterly direction, while Source 2 moves in a south-westerly direction. These propagation directions are coincident with the expansion of the associated CME and its shock wave outwards through the corona. However, from Earth's view, this plane-of-sky movement is affected by significant line-of-sight projection effects. Therefore, we construct a 3D perspective using observations from STEREO that were at ideal locations to provide side views of the eruption (see insert in Fig.~\ref{fig:fig4}a). The STEREO images in Fig.~\ref{fig:fig4}b--c are composite images of the CME observed in the low corona at EUV wavelengths observed by the Extreme Ultraviolet Imager (EUVI) and white-light images from the COR1 coronagraph that observed the CME higher in the corona (the COR1 field of view extends between 1.5 and $4\,R_{\odot}$). The CME expands radially and outwards from the solar limb in both STEREO-A and -B fields of view. }

{To determine the position of the radio sources in relation to the CME in the STEREO perspectives, we de-project the radio source centroids from their plane-of-sky view. Assuming that the radio bursts are emitted at the fundamental plasma frequency and do not originate from accelerated electrons inside the CME (see previous section), it is possible to determine their approximate height in the SDO perspective. The height of the radio bursts on the images away from the plane of sky is generally unknown, however electron density models of the solar corona can be used to provide the density stratification as a function of height to de-project the radio source locations \citep{mo20}. In the case of plasma emission, the electron density ($n_e$) can be related to the plasma frequency ($f_p$) using the following: $f_p \sim 9000\sqrt{n_e}$. Therefore, for a certain radio burst emitted by the plasma emission mechanism, we can estimate its radial distance from the solar surface. We use the radially-symmetric electron density model of \citet{new61} to estimate the heights of the moving radio sources at a specific frequency. Assuming the radio bursts are emitted at the fundamental plasma frequency, the radial distance corresponding to the plasma frequencies of 150 and 173~MHz, respectively, is calculated using the four-fold Newkirk density model. This model also accounts for enhanced densities above active regions. This distance, combined with the plane-of-sky coordinates of the centroids, is then used to project the coordinates of the radio sources onto the STEREO planes and to represent them in 3D. There is however an uncertainty in estimating the radial distances of radio bursts using density models. Firstly, the solar corona is variable and other density models can predict different distances. We have therefore assigned an uncertainty of $\pm0.3\,R_{\odot}$ to the height estimates of the four-fold Newkirk model to reflect a range of possible heights of the radio sources. This was based on two other density models, one of the background solar corona \citep{sa77} and one of the more active solar corona achieved through the six-fold Newkirk model. We also assign an uncertainty in estimating the plane-of-sky NRH centroids of $\pm0.1\,R_{\odot}$, which corresponds approximately to half of the extent of the full width at half maximum of the radio sources.}

{Using the 3D coordinates of the centroids, their approximate locations can be projected on the STEREO-B perspective for Sources 1 and 3 and on the STEREO-A perspective in the case of Source 2 (Fig.~\ref{fig:fig5}). The projections onto the STEREO planes reveal where the radio sources are located in relation to the CME and in which direction they move in relation to the CME expansion. All moving sources appear to be coincident with the southern CME flanks and move laterally in concert with the lateral expansion of the CME in the STEREO perspectives. In the case of Sources 1 and 2, the centroids are located outside the CME, in a region where the CME expands into a coronal streamer (labelled in Fig.~\ref{fig:fig5}). Source 3 appears to be located low inside the CME in the STEREO-B projection, however still closely related to the southern CME flank, and in 3D this source could be located outside, close to the CME legs. Despite the large uncertainty associated with such a de-projection technique, the used method would not affect the kinematics of the moving radio bursts in the STEREO perspectives, and these radio bursts would still follow the CME flank expansion to the south. We note that changing the density model used to de-project the centroids also does not affect the kinematics and position of radio sources at the CME flanks, the only effect being that the centroid position shifts either closer or farther away with respect to the flank.}

{At the southern CME flanks, where the moving radio bursts are located, the CME encounters a coronal streamer indicated in Fig.~\ref{fig:fig5}. The streamer appears to be perturbed by the passage of the CME, since its outline is changing in the running-difference images shown in Fig.~\ref{fig:fig5}. Since the streamer is located outside the CME, a wave driven by the CME expansion would most likely be able to perturb it. Densities inside coronal streamers are considered to be high, therefore the Alfv{\'e}n speed inside streamers is low \citep{ev08}. Furthermore, the CME shows fast lateral expansion in the LASCO images in Fig.~\ref{fig:fig1} and in the STEREO images in Fig.~\ref{fig:fig5}.}

\begin{figure}[ht]
\centering
    \includegraphics[width=0.95\columnwidth, trim = {0px 40px 20px 40px}, clip]{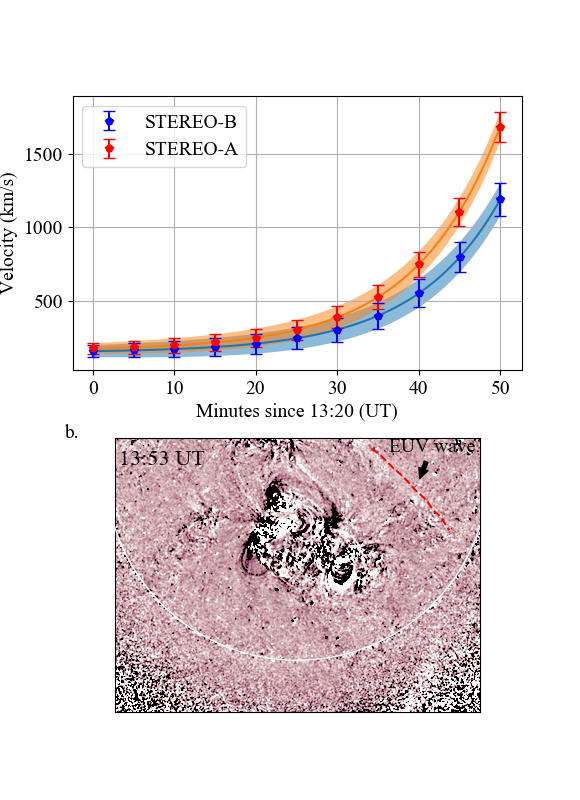}
    \caption{(a) Lateral CME speed along the southern flank as observed in STEREO-A and -B. The CME speed data points were extracted from the composite EUVI and COR1 images from 13:20 through 14:10~UT, at 5-minute intervals. The CME reaches a speed of >1000~km/s at 14:10~UT in both STEREO-A and -B. (b) Faint EUV wave associated with the CME eruption, observed at the time of Sources 2 and 3. The EUV wave was only observed on the western hemisphere. The fast CME lateral speed after 13:55~UT (>500~km/s) and the presence of an EUV wave around the same time are strong indicators of a propagating shock wave surrounding the CME during the occurrence of Sources 2 and 3.}
    \label{fig:fig6}
\end{figure}

\begin{figure*}[ht]
\centering
    \includegraphics[width=0.8\linewidth, angle = 0]{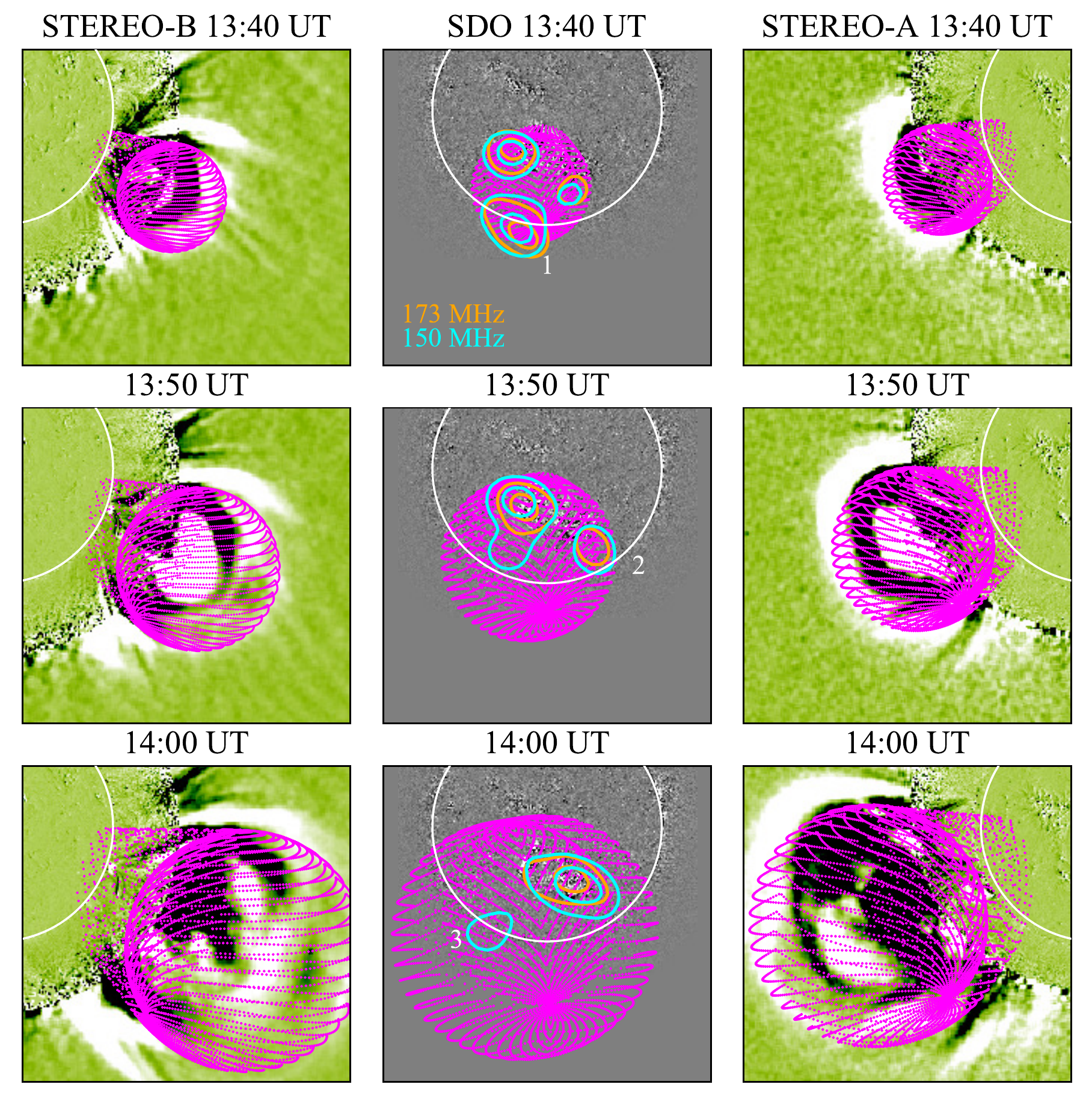}
    \caption{3D fitting of the CME bubble from three perspectives. The perspectives are images from STEREO-B, SDO, and STEREO-A from left to right, at three different times from top to bottom (13:40, 13:50, and 14:00~UT). The perspectives show the expansion of the CME fitted with the GCS model applied on the CME bubble (magenta dots). The radio contours at two frequencies (150 and 173~MHz) are overlaid in the middle panels on the SDO images. }
    \label{fig:fig7}
\end{figure*}

{The CME speed was estimated by fitting an exponential height--time function, $h(t)$ from \citet{ga03}, to plane-of-sky EUVI and COR1 images from STEREO-A and -B. The lateral expansion speed of the CME was estimated from 13:20 to 14:10~UT by tracking the southern flank of the expanding CME in both STEREO-A and -B. Since the CME is located at the limb in STEREO-A and -B, we expect an accurate estimate of the lateral expansion of the CME from this perspective. The height--time function has the following form:
\begin{equation}
h(t) = h_0 + v_0t + a_0 \tau^2 \mathrm{exp}(t/\tau) \, ,
\end{equation}
where $h_0, v_0$, and $a_0$ are the initial height, velocity and, acceleration, respectively. The function from \citet{ga03} fitted to the CME expansion contains an exponentially varying acceleration of the form $a_0 \mathrm{exp}(t/\tau)$. The CME was found to expand rapidly towards the southern pole, reaching a speed of >1000~km/s at 14:10~UT (Fig.~\ref{fig:fig6}a). This speed is larger than Alfv{\'e}n speed values predicted inside coronal streamers of up to ${\sim}400$~km/s \citep{ev08}. The wave perturbing the streamer could in fact be a shock wave driven by the CME \citep{ch10, fe12}, especially after 13:55~UT when the CME speed is >500~km/s (during the onset of Sources 2 and 3). A white-light enhancement indicating a shock is indeed observed surrounding the CME in LASCO/C2 images at a later time (Fig.~\ref{fig:fig1}c).}

{More evidence of a low-coronal shock is provided by the presence of a faint coronal or EUV wave (Fig.~\ref{fig:fig6}b). Running-difference images from SDO/AIA at 211~{\AA} show a faint propagating structure between 13:50 and 14:00~UT propagating through the western hemisphere. This structure resembles a faint EUV wave, which represents a disturbance of the low-coronal plasma caused by the outward expansion of the CME \citep{lo08}. The shock wave driven by the CME in the low corona is believed to manifest itself as the EUV wave across the solar disc \citep{as09,do12}, therefore CME shocks, EUV waves, and particle acceleration are closely related \citep{ko11, gr11}.} 



\subsection{3D reconstruction of the electron acceleration locations}

{The three perspectives from SDO--SOHO and the two STEREO spacecraft allow us to reconstruct the 3D structure of the CME evolving through the solar corona. Namely, we use the graduated cylindrical shell \citep[GCS;][]{the06,the09} model, which consists of a parameterised croissant-shaped grid, to manually trace the CME extent through time in the different planes of sky. The 3D reconstruction of the CME is then performed by overlaying the parametrised grid onto the three planes of sky (SOHO, STEREO-A, and STEREO-B) to reproduce the observed features at its best. We perform reconstructions from 13:40 to 14:10~UT (STEREO time) at five-minute intervals. Since the CME first appeared in the LASCO/C2 field of view at 14:12~UT, we perform reconstructions using only the COR1-A and -B viewpoints from 13:40 to 14:05~UT. We note that the fit to the CME bubble includes part of the nearest deflected streamer at the southern flank (visible in STEREO-A images in the middle panels of Fig.~\ref{fig:fig5}), as this is indicative of a shock wave perturbing the environment around the CME.}

\begin{figure*}[ht]
\centering
    \includegraphics[width=0.57\linewidth, trim = {50px 10px 70px 10px}, clip, angle = -90]{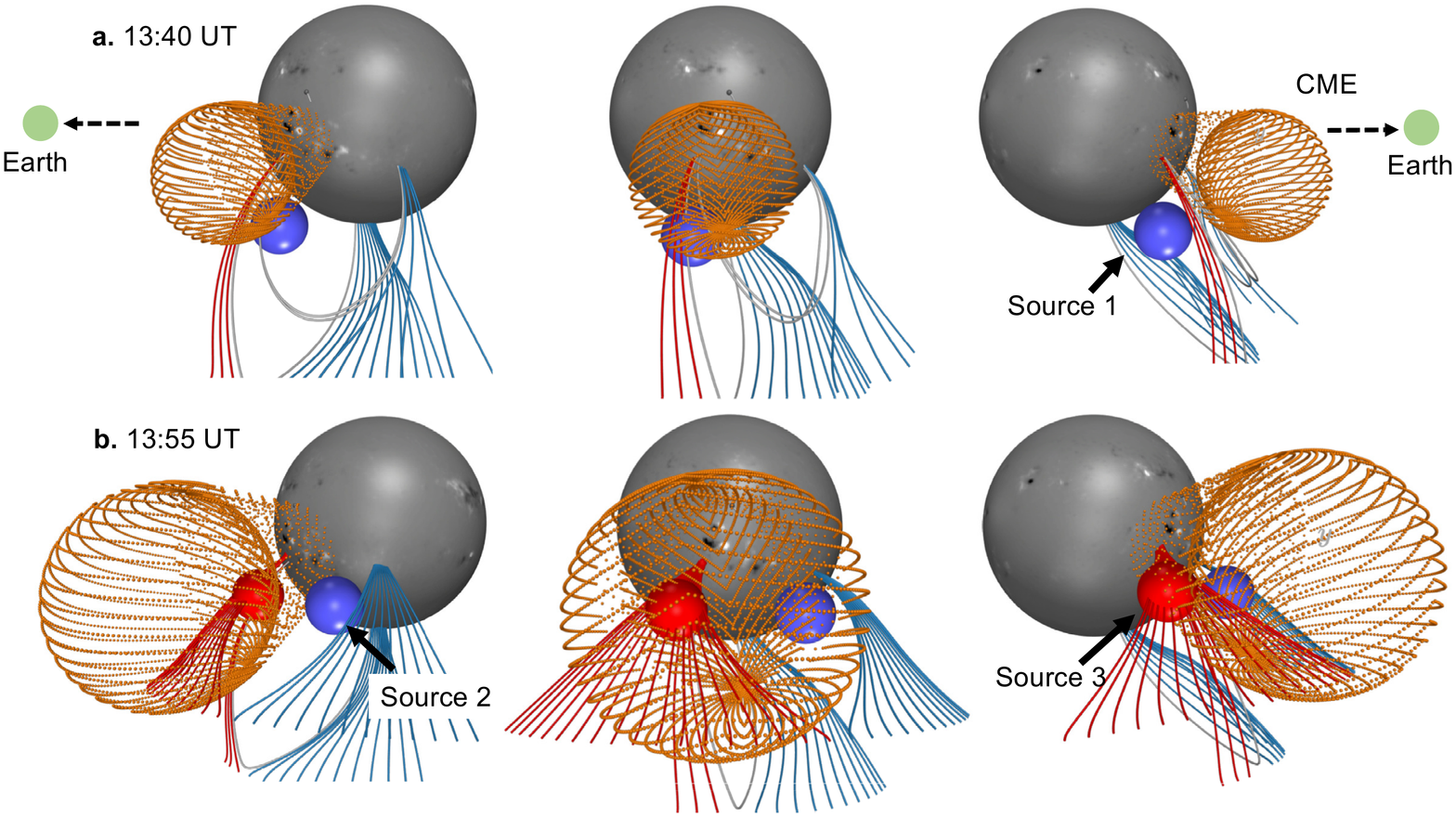}
    \caption{3D model of the CME bubble and electron acceleration locations. Each panel in this figure shows a different perspective similar to that of STEREO-A (left), plane-of-sky such as the SDO plane (middle), and STEREO-B (right). The magnetic field polarity (white for positive and black for negative) from photospheric magnetograms is overlaid on top of the surface of the solar sphere. (a) Reconstruction of the CME eruption in the solar corona at 13:40~UT during the occurrence of Source 1. The orange mesh represents the CME bubble and the extrapolated magnetic field lines around the CME are plotted as red (negative polarity) and blue (positive polarity) lines for the open field and grey lines for the closed field. Source 1 (blue sphere) is located outside the CME. (b) Reconstruction of the CME eruption in the solar corona at 13:55~UT during the occurrence of Sources 2 and 3, which are located on either side of the CME. These radio sources are also located in open field regions surrounding the CME. }
\label{fig:fig8}
\end{figure*}

{The reconstructed 3D CME bubble (wireframe consisting of magenta dots) is shown at three different times from top to bottom (13:40, 13:50, and 14:00~UT) and from the three perspectives from left to right (STEREO-B, SDO, and STEREO-A) in Fig.~\ref{fig:fig7}. The middle panels of Fig.~\ref{fig:fig7} show that the CME expanded very rapidly laterally in the plane of sky from Earth's view within a 30-minute period. The speed of the CME southern flank from STEREO-B observations is also high, up to 1000~km/s before 14:10~UT (as determined in the previous section). The radio source contours (blue contours at 150~MHz and orange contours at 170~MHz) are overlaid on the SDO images in the middle panels of Fig.~\ref{fig:fig7}. The contours of the first moving source at 13:40~UT (at 30\% and 70\% of the maximum intensity level) are located slightly outside the CME bubble. The radio contours of the second and third moving radio bursts appear to be located inside the CME bubble in the SDO viewpoint in Fig.~\ref{fig:fig7}. However, these radio sources can in fact be located behind the expanded CME bubble based on the centroid locations in Fig.~\ref{fig:fig5} and on either side of the CME flanks. In the STEREO-A and -B images in Fig.~\ref{fig:fig5} the radio centroids are indeed located at the CME flanks, which correspond to locations on either side of the CME legs in the SDO plane.}

\begin{figure}[ht]
\centering
    \includegraphics[width=0.74\columnwidth, trim = {30px 50px 80px 70px}, clip, angle = -90]{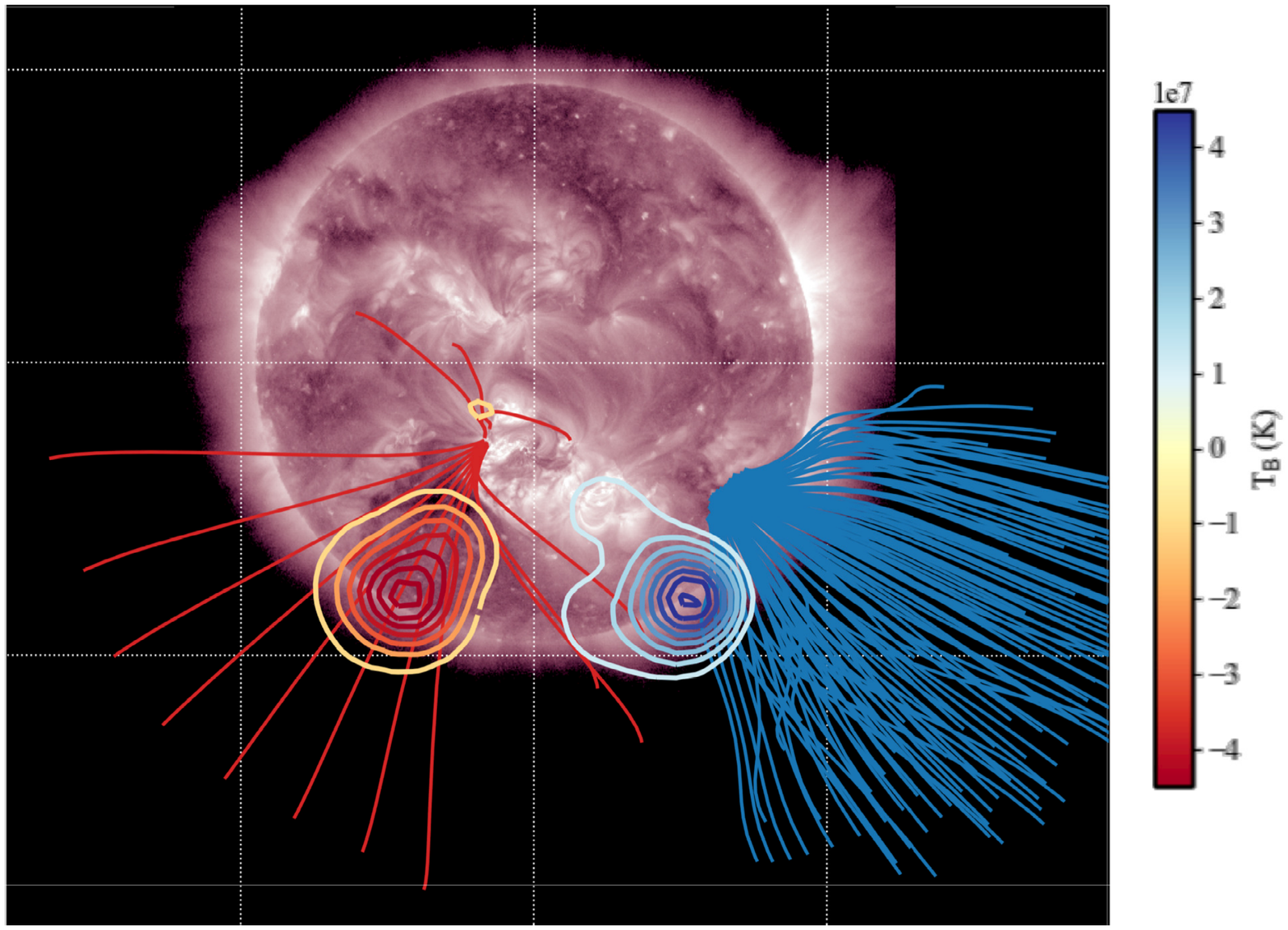}
    \caption{Polarised moving radio bursts and open magnetic field regions. Radio contours and extrapolated open magnetic field lines are overlaid on AIA 211~{\AA} images at 13:50~UT. The positive open field lines (blue) and left-handed circularly polarised burst (blue contours) are located to the right of the active region where the CME originates, while the negative open field lines (red) and right-handed circularly polarised burst (red contours) are located to the left of the active region. The sense of polarisation is from the point of view of the observer. The two radio sources have an opposite sense of circular polarisation since they propagate on either sides of the active region, through open field regions also of opposite polarity.}
    \label{fig:fig9}
\end{figure}

{The CME reconstruction can be used to visualise the eruption and associated accelerated particles in 3D. The panels of Fig.~\ref{fig:fig8} show the CME bubble and the coronal magnetic field, together with the location of the radio source centroids in 3D. In each of the panels in Fig.~\ref{fig:fig8}, the orange mesh represents the shape of the CME bubble in 3D and the extrapolated magnetic field lines around the CME are plotted as red and blue lines for the open field and grey lines for the closed field. Source 1 at 13:40~UT (blue sphere in the top row) is clearly located outside the CME at the southern flank. Source 2 (blue sphere in the bottom row) and Source 3 (red sphere in the bottom row) at 13:55~UT are located on either side of the CME western and eastern flanks, respectively. Based on the 3D reconstruction, the moving radio bursts observed in the plane of sky (the LASCO and SDO planes) are therefore the electromagnetic radiation produced at the CME flanks, most likely by shock-accelerated electrons given the fast speed of the CME expansion. These electrons are accelerated behind the overlying CME bubble seen in the SDO perspective, but our reconstruction shows that they can be located outside the CME bubble in other perspectives.}


\subsection{Magnetic environment of the accelerated electrons}

{Another noteworthy finding of the moving radio bursts studied here is their sense of circular polarisation. Langmuir (plasma) waves are generated by accelerated electron beams propagating through a plasma. For Langmuir waves travelling parallel or anti-parallel to the magnetic field, the polarisation is in the sense of the ordinary (o-) mode \citep{du76, du80}, which is the only mode that can propagate through a coronal plasma to produce fundamental plasma emission \citep{me09}. Fundamental radiation is then expected to be up to 100\% polarised \citep{me72}. The o-mode resides on the same dispersion surface as the l-mode (the left-handed circularly polarised mode) and, hence, o-mode electromagnetic waves become circularly polarised following the refraction of the waves to the field-aligned direction \citep{oa79}.}

{The polarisation of the moving radio bursts in this study, emitted at the fundamental plasma frequency, is most likely also in the sense of the o-mode. In particular, Sources 2 and 3 have a high degree of circular polarisation of up to 90\% and, at the same time, an opposite (left and right, respectively) sense of circular polarisation from the viewpoint of the observer, which remains consistent throughout their duration. Extrapolations of the photospheric magnetic field \citep{sc03} show that the CME is surrounded on either side by open-field regions of opposite polarity, in the plane of sky (Fig.~\ref{fig:fig9}). Since Sources 2 and 3 are also located on either side of the CME in the plane of sky, they are emitted in these open field regions of opposite polarity, where red field lines denote magnetic field from negative polarity regions and blue field lines denote magnetic field from positive polarity regions (Fig.~\ref{fig:fig9}). The magnetic field is oppositely oriented in the case of these two radio sources, therefore, their sense of polarisation is also opposite from the point of view of the observer. }

{Opposite senses of circular polarisation at low radio frequencies have been reported before by \citet{kai70} in the case of type I noise storms (stationary emission accompanying active regions) and stationary type IV radio bursts, both believed to originate due to electrons accelerated inside small-scale coronal loops. Moving type IV radio bursts with opposite senses of circular polarisation have also been previously observed by \citet{st71}. However, these radio sources initially had a lower degree of circular polarisation (30\%), and one of the moving sources lasted for over an hour, most likely representing synchrotron radiation from mildly relativistic electrons as concluded in \citet{st71}. One of the radio sources observed by \citet{st71} became up to 100\% circularly polarised before it faded, while \citet{sm71} also presented previous observations of moving radio sources with a very high degree of circular polarisation, similar to our observations. A high degree of circular polarisation (>50\%) has also been identified in early observations of herringbones \citep{su80}. Similar moving radio bursts have also been observed by \citet{sm71}, however in the case of opposite polarity type IVs an expanding arch structure was identified, believed to be generated by synchrotron electrons. More recently, opposite senses of circular polarisation have also been observed in the case of opposite-polarity neighbouring coronal holes \citep{mc19}. Similarly to our findings, \citet{ka78} also concluded that the polarity of magnetic fields at the type IV emission sites is similar to the polarity of magnetic fields at the photosphere based on a statistical study of 30 moving type IV bursts. \citet{ka78} showed that left-handed polarised bursts are predominantly associated with positive polarity fields and right-handed bursts are the opposite, as in Fig.~\ref{fig:fig9}. }

\subsection{Propagation effects on the 3D position of radio sources}

{Radio emission produced close to the plasma frequency level can be affected by propagation effects \citep{ro83, po88, mc18}. In our case, the radio sources emitted at the fundamental plasma frequency are therefore emitted close to the plasma level and are likely to be impacted by propagation effects such as refraction and scattering by density inhomogeneities in the corona. These effects can shift the true position of the observed radio source \citep{ro83}, often resulting in radio sources at higher altitudes than the predicted plasma level \citep{po88, mc18}. However, \citet{mc18} argued that propagation effects only partially affect the position of the radio emission observed and that is is possible that the electrons responsible for this emission travel through over dense structures.  }

{In our study, we used a four-fold Newkirk model to account for higher heights where low-frequency radio emission is observed. If propagation effects are indeed significant, our density model may predict too high densities to accurately reconstruct the radio emission in 3D. However, the uncertainties in the radio source centroids in Fig.~\ref{fig:fig5} were determined based on the positions predicted by a background density model of the solar corona on the lower end. A background model predicts that the source positions are closer to the CME and shifted in the direction of the CME core. These locations would place Source 2 completely inside the reconstructed CME shell, while Source 3 is partially inside the reconstructed shell within this uncertainty. Source 1 remains unaffected within this uncertainty. Therefore, if propagation effects are significant or the radial density model used is not adequate, there is a possibility that Sources 2 and 3 represent plasma radiation from inside the CME. We note, however, that this would have no effect on the kinematics of the moving radio bursts. All of the radio sources would still be closely related to the CME lateral expansion in the low corona, closely following the propagation of the CME flanks. }

\begin{figure*}[ht]
\centering
    \includegraphics[width=0.47\linewidth, trim = {70px 20px 120px 10px}, clip, angle = -90]{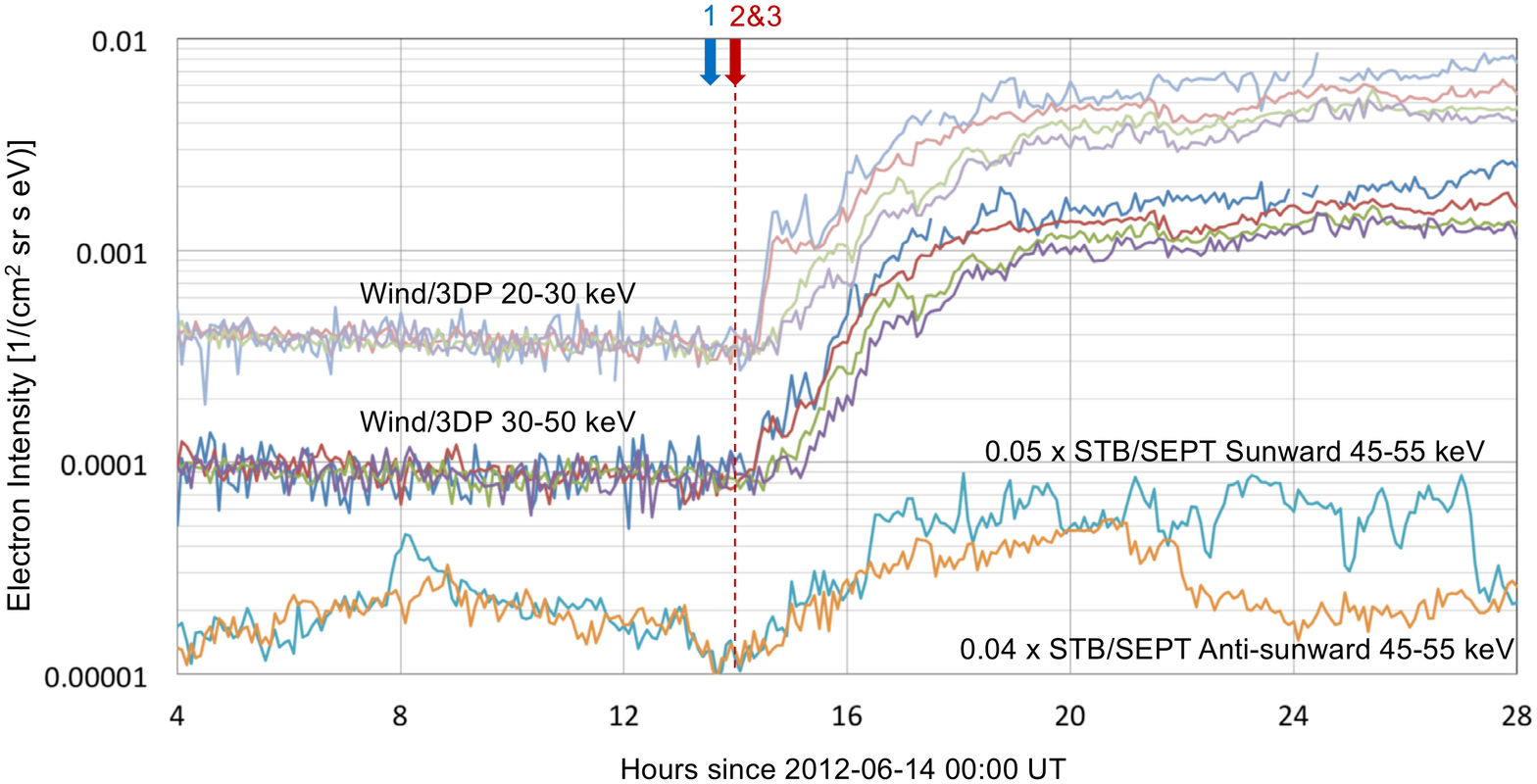}
    \caption{Energetic electrons observed in situ at 1~AU. The electrons were observed by two spacecraft: STEREO-B/SEPT at 45--55~keV and WIND/3DP at 20--30 and 30--50~keV. The start time of Sources 1, 2, and 3 is annotated with the blue and red arrows. The STEREO-B electron intensities were multiplied by factors of 0.05 for the sunward-looking telescope and by 0.04 for anti-sunward-looking telescopes and the data were normalised so that the intensities agree at 16:00 and 20:00~UT. The electron intensities observed by the two spacecraft started increasing shortly after the onset of Sources 2 and 3, denoted by the vertical red dashed line.}
    \label{fig:fig10}
\end{figure*}


\section{Discussion} \label{sec:discussion}

{We observed three moving radio sources with a high degree of circular polarisation that are best interpreted as plasma radiation. The nature and origin of the radio sources are interpreted as follows:}

{Source 1 occurs during the early stages of the CME eruption. In plane-of-sky observations, Source 1 appears to be located inside the apex of the CME, however in 3D reconstructions, the radio source is located significantly outside of the CME at its southern flank. There is however a discrepancy between the available orientation of open field and the sense of circular polarisation of Source 1. Source 1 has the same sense of circular polarisation as Source 2, however it occurs at a similar location as Source 3, where the orientation of the open field is negative. It is possible that we need to revise our interpretation of Source 1 as plasma emission outside the CME. Source 1 is unlikely to be related to a CME flank shock, as the CME may not be fast enough to drive a shock at this time (the lateral CME speed is <500~km/s before 13:50~UT when Source 1 occurs). Source 1 is also less polarised than Sources 2 and 3 and has less steep spectral indices that fall in the range of gyro-synchrotron spectral indices \citep{ni04}. However, this emission appears narrowband in the NRH images and, given the lack of imaging at lower frequencies, it is not possible to draw other conclusions on the nature of this emission. Another possibility is that Source 1, that occurs at an earlier time, has access to differently oriented closed magnetic field lines compared to Source 3, to explain its sense of circular polarisation. }

{In the case of Sources 2 and 3, we provided evidence that they are emitted by plasma radiation that moves outwards with the CME flanks. In the case that scattering is not significant at NRH frequencies, the emission is predominantly located outside the reconstructed CME shell. At this time, we also showed that a faint coronal wave is present on the western hemisphere and that the CME becomes fast enough to drive a low coronal shock. A CME-driven shock is a likely possibility for the acceleration of energetic electrons producing radio emission outside the CME. In the event that scattering is indeed significant at the NRH frequencies, then these radio sources, in particular Source 3 would be located inside the reconstructed CME shell. In this case, the radio emission would represent plasma radiation inside the CME legs and the origin of accelerated electrons may be a reconnecting current sheet behind the CME \citep{de12, mo19b}. Thus, the radio sources would also move outwards along a constant density gradient with the CME, however they would not be located at the flank but inside the CME behind its flanks. In either case, the radio source propagation is in the direction of the CME lateral expansion and the moving radio sources are closely related to the propagation of the CME flanks through the low corona. } 

{The availability of open magnetic field lines and regions of high density plasma inside the coronal streamers surrounding the CME appears to facilitate the acceleration of electrons at the CME flanks. The presence of open field lines can also facilitate the escape of energetic electrons to interplanetary space. Energetic electrons associated with this CME have indeed been observed in situ at 1~AU (astronomical unit, where 1~AU = $1.5\times10^{11}$~m ) by the Solar Electron and Proton Telescope \citep[SEPT;][]{lu08} on board STEREO-B and the Three-Dimensional Plasma \citep[3DP;][]{li95} instrument on board Wind \citep{og97} located near Earth. Both spacecraft observed an increase in electron intensities shortly after the onset of Sources 2 and 3 (Fig.~\ref{fig:fig10}). The electrons observed by Wind in two energy channels (lighter shades for 20--30~keV and darker shades for 30--50~keV in Fig.~\ref{fig:fig10}) have pitch angles of 11$^{\circ}$ (blue), 56$^{\circ}$ (red), 101$^{\circ}$ (green), and 146$^{\circ}$ (purple), indicating that they are travelling from a sunward direction. The near-relativistic electrons observed by STEREO-B are plotted from the sunward (blue) and anti-sunward (orange) directions and have energies of 45--55~keV in Fig.~\ref{fig:fig10}. }

{The presence of in-situ electrons indicates a low-coronal acceleration region co-temporal with the appearance of Sources 2 and 3. Assuming a travel path of 1.2~AU for the electrons, and a time of 8~minutes for the radio emission to travel from the Sun to Earth, the delay times of the energetic electrons with respect to Sources 2 and 3 are the following: 15.9~min at 50~keV, 22.1~min at 30~keV, and 28.4~min at 20~keV. Accounting for these delay times, the acceleration of the in-situ electrons in the corona occurs shortly or only a few minutes after the onset time of Sources 2 and 3. Both radio sources and energetic electrons start at a later stage in the eruption and not during the impulsive energy release of the flare. In the case of in-situ electrons, this indicates the presence of a low-coronal shock wave accelerating these particles \citep{po11}. Sources 2 and 3 are short-lived in comparison to the electron event; however, as the CME continues to propagate further away from the Sun, any possible radio emission would move outside the NRH frequency band towards lower and lower frequencies. The coincidental onset time of the in-situ electrons and Sources 2 and 3 may indicate that both events originated at the CME flanks, and a CME shock may be the origin of the observed in-situ electrons. In addition, Sources 2 and 3 occurred in regions where the CME had ample access to open magnetic field lines, where electrons could escape along the interplanetary magnetic field. The regions where Sources 2 and 3 originate may therefore be related to the origin of the observed in-situ energetic electrons.}


\section{Conclusions} \label{sec:conclusions}

{So far, moving radio bursts have been classified either as emission due to electrons trapped inside the CME or electrons accelerated by magnetic reconnection processes outside the CME in the case of type IVs. However, moving radio bursts coincident with type IV emission and CME shocks have not been classified before, unless they show clear type II or herringbone structures. Here, we have shown a new origin of moving radio sources, other than type IIs or herringbones, that are generated at the CME flanks. Some of the moving bursts are likely associated with electrons accelerated by the CME-driven shock or electrons accelerated inside the CME also moving outwards with the CME flanks. The moving radio bursts in this study behave similarly to type II radio bursts or herringbones propagating outwards in the solar corona \citep{ca13, zu18, mo19a}. Simulations of type II radio bursts, emitted during CMEs with strong lateral expansions such as the one observed here, can produce broad fundamental and harmonic type II lanes in dynamic spectra, that are indistinguishable from each other \citep{kn05}. This could be the case in our observations, where we do not observe any type II emission lanes in dynamic spectra but instead a broader continuum emission that resembles a type IV with superimposed fine structures.}

{In our study, plane-of-sky ground observations were not sufficient to identify the nature of the moving radio emission observed. The additional perspectives provided by the STEREO mission were invaluable for this study. Future observations of moving radio bursts will benefit from better imaging capabilities at multiple frequencies combined with high-resolution spectroscopy to be able to distinguish the type of radio emission observed. Such observations are now possible with the Low Frequency Array (LOFAR) or the Murchinson Widefield Array (MWA). These observations, combined with additional vantage points from the STEREO-A spacecraft and future L5 missions, can provide the opportunity to understand the 3D nature and propagation of CMEs and associated radio emission}.


\begin{acknowledgements}{The results presented here have been achieved under the framework of the Finnish Centre of Excellence in Research of Sustainable Space (Academy of Finland grant numbers 312390, 312357, and 312351), which we gratefully acknowledge. E.P. acknowledges the Doctoral Programme in Particle Physics and Universe Sciences (PAPU) at the University of Helsinki. R.V. acknowledges the financial support of the Academy of Finland (project 309939). M.P. acknowledges the European Research Council (ERC) Consolidator grant 682068-PRESTISSIMO, and Academy of Finland grants 312351 and 309937. E.K.J.K. acknowledges the ERC under the European Union's Horizon 2020 Research and Innovation Programme Project SolMAG 724391, and Academy of Finland Project 310445. We would like to acknowledge the Nan{\c c}ay Radioheliograph, funded by the French Ministry of Education and the R\'egion Centre in France, the Nan{\c c}ay Decametric Array hosted at the  Nan{\c c}ay Radio Observatory/Unit\'e Scientifique de Nan{\c c}ay of the Observatoire de Paris (USR 704-CNRS, supported by Universit\'e de Orl\'eans, OSUC and the R\'egion Centre) and the e-Callisto Birr spectrometer located at the Rosse Observatory and supported by Trinity College Dublin. We thank the Radio Solar Database service at LESIA \& USN (Observatoire de Paris) for making the NRH/NDA data available. We would like to thank the Wind/3DP and STEREO/SEPT instrument teams for making their data publicly available. }\end{acknowledgements}

\bibliographystyle{aa} 
\bibliography{aanda.bib} 

\end{document}